\def\nuebar{\bar{\nu}_e}
\def\numubar{\bar{\nu}_\mu}
\def\mev{\,{\rm MeV}}
\def\misec{\,\mu{\rm s}}
\def\today{April 21, 1995}
\documentstyle[preprint,aps]{revtex}
\begin{document}
\draft
\title{Candidate Events in a Search for $\bar \nu_{\mu} \rightarrow
\bar \nu_e$ Oscillations}
\author{C. Athanassopoulos$^{12}$, L. B. Auerbach$^{12}$,
R. Bolton$^7$, B. Boyd$^9$, R. L. Burman$^7$,
\\D. O. Caldwell$^3$, I. Cohen$^6$, J. B. Donahue$^7$,
A. M. Eisner$^4$, A. Fazely$^{11}$,
\\F. J. Federspiel$^7$, G. T. Garvey$^7$, M. Gray$^3$, R. M.
Gunasingha$^8$,
V. Highland$^{12 \:\dag}$, \\R. Imlay$^8$, K. Johnston$^{9}$,
W. C. Louis$^7$,
A. Lu$^3$,
J. Margulies$^{12}$,  \\K. McIlhany$^{1}$, W. Metcalf$^8$,
R. A. Reeder$^{10}$, V. Sandberg$^7$,
M. Schillaci$^7$, \\D. Smith$^{5}$, I. Stancu$^{1}$, W.
Strossman$^{1}$,
G. J. VanDalen$^{1}$,
W. Vernon$^{2,4}$, \\Y-X. Wang$^4$, D. H. White$^7$, D. Whitehouse$^7$,
D. Works$^{12}$, Y. Xiao$^{12}$}
\address{$^{1}$University of California, Riverside, CA 92521}
\address{$^{2}$University of California, San Diego, CA 92093}
\address{$^3$University of California, Santa Barbara, CA 93106}
\address{$^4$University of California
Intercampus Institute for Research at Particle Accelerators,
Stanford, CA 94309}
\address{$^{5}$Embry Riddle Aeronautical University, Prescott, AZ
86301}
\address{$^6$Linfield College, McMinnville, OR 97128}
\address{$^7$Los Alamos National Laboratory, Los Alamos, NM 87545}
\address{$^8$Louisiana State University, Baton Rouge, LA 70803}
\address{$^{9}$Louisiana Tech University, Ruston, LA 71272}
\address{$^{10}$University of New Mexico, Albuquerque, NM 87131}
\address{$^{11}$Southern University, Baton Rouge, LA 70813}
\address{$^{12}$Temple University, Philadelphia, PA 19122}
\address{$^{\dag}$ deceased}

\date{\today}
\maketitle

\vfill\eject
\begin{abstract}
    A search for $\nuebar$'s in excess of the number expected from
conventional sources has been made using the Liquid Scintillator
Neutrino Detector, located 30 m from a proton beam dump
at LAMPF.  A $\nuebar$ signal was detected via the reaction
$\nuebar\,p \rightarrow e^{+}\,n$ with $e^+$ energy between 36 and
$60\mev$, followed by a $\gamma$ from $np\rightarrow d\gamma$
($2.2\mev$).
Using strict cuts to identify
$\gamma$'s correlated with positrons
results in a signal of 9 events, with an expected background
of $2.1 \pm 0.3$.  A likelihood fit to the entire $e^+$ sample
yields a total excess of
$16.4^{+9.7}_{-8.9}\pm 3.3$
events, where the second uncertainty
is systematic.  If this excess is attributed to neutrino
oscillations of the type $\numubar\rightarrow\nuebar$, it corresponds
to an oscillation probability of
($0.34^{+0.20}_{-0.18}\pm 0.07$)\%.
\end{abstract}


Neutrino mass is a central issue for particle physics,
because neutrinos are massless in the Standard Model, and for
cosmology,
because the relic neutrinos, if massive, would have profound
effects on the structure of the universe.
To search for such mass an experiment has been carried out
using neutrinos from $\pi$ and $\mu$ decay at rest from the
Los Alamos Meson Facility (LAMPF) beam stop.
Observation of $\bar \nu_e$ production above that expected from
conventional
processes may be interpreted as evidence for $\bar \nu_{\mu}
\rightarrow
\bar \nu_e$ oscillations (and hence mass) or some direct lepton number
violating process.

Protons from the LAMPF 800-MeV linac produce pions in a 30-cm-long
water
target positioned approximately 1 m upstream from the copper beam stop.
\cite{LSNDref}
The beam stop provides a source of $\numubar$, via
$\pi^+ \rightarrow \mu^+ \nu_{\mu}$ followed by
$\mu^+ \rightarrow e^+ \nu_e \numubar$ decay-at-rest; the relative
$\nuebar$ yield is $\sim 4 \times 10^{-4}$ \cite{nuebkg} for
$E_{\nu}>36 \mev$.
The Liquid Scintillator Neutrino Detector (LSND)
detects $\nuebar$ by $\nuebar p \rightarrow e^+ n$,
followed by a $\gamma$ from $np \rightarrow d\gamma$ (2.2 MeV).
Requiring an electron energy above $36\mev$ eliminates most of the
accidental background from
$\nu_e ^{12}C \rightarrow e^- X$, while the upper energy requirement of
$60\mev$ allows for
the $\numubar$ endpoint plus energy resolution.
The 7691 coulombs of protons
were obtained in a 1.5-month run in 1993
and a 3.5-month run in 1994.
The calculated $\bar \nu_{\mu}$ decay-at-rest flux
\cite{beam_mc} totaled $3.75 \times 10^{13}
\nu/\rm cm^2$ at the center of the tank, with an uncertainty of 7\%.

The center of the detector is 30 m
from the neutrino source and is shielded by the equivalent
of 9 m of steel.
The detector, an approximately cylindrical tank 8.3 m long by
5.7 m in diameter, is under 2kg/cm$^2$ of overburden to reduce the
cosmic-ray flux and is located at an angle of $12^o$ relative
to the proton beam direction.
On the inside surface
of the tank 1220 8-inch Hamamatsu phototubes provide $25\%$
photocathode coverage with uniform spacing. The tank is filled with
167 metric tons of liquid scintillator consisting of mineral oil and
0.031 g/l
of b-PBD. The composition of the liquid is $CH_2$, including $1.1\%$
of $~^{13}C$ and $\sim 10^{-4}$ of $~^{2}H$.
The low scintillator concentration allows the detection of both
\v{C}erenkov light and scintillation light and yields an
attenuation length of more than 20 m for
wavelengths greater than 400 nm.
A sample
of $\sim 10^6$ electrons from cosmic-ray muon decays
in the tank was used to
determine the electron energy scale and resolution.
A typical electron at the end-point energy
of 52.8 MeV
leads to $\sim 1750$ photoelectrons, of which $\sim 300$
are in the \v{C}erenkov cone.
The phototube
time and pulse height signals are used to reconstruct the electron
track
with an average r.m.s. position
resolution of $\sim 30$ cm, an angular resolution of $\sim 12^o$,
and an energy resolution of $\sim 7\%$.
A liquid-scintillator
veto shield \cite{VETO} surrounds the detector tank
with 292 5-inch phototubes.

Particle identification (PID) for relativistic particles is based upon
the \v{C}erenkov cone
and the time distribution of the
light,\cite{Reeder} which is broader for non-relativistic particles.
Three PID quantities are used:
the \v{C}erenkov cone fit quality, the
event position fit quality, and the fraction of phototubes hit at a
time
corresponding to light emitted more than 12 ns later
than the reconstructed event time.
Comparing electrons from cosmic-ray muon decays with
cosmic-ray-produced neutrons of similar deposited energy,
a neutron rejection
of $\sim 10^{-3}$ is achieved with an electron efficiency of $79\%$.

Each phototube channel is digitized every 100ns and the
data is stored in a circular buffer.
A primary event
trigger is generated when the total number of hit phototubes in two
consecutive 100 ns periods
exceeds 100. However, no primary
triggers are allowed for a period of $15.2 \mu$s
following veto shield events
with $> 5$ hit veto phototubes in order to reject electrons
from the decay of stopped cosmic-ray muons in the detector.
The trigger operates independently of the state of the
proton beam,
so the beam duty factor of 7.3\% allows 13 times more
beam-off than beam-on data to be collected.
After a primary trigger with $>125$ hit phototubes ($>300$ in 1993),
the
threshold is lowered to 21 hit phototubes for a period of 1 ms
in order to record the
 2.2 MeV $\gamma$ from $np \rightarrow d\gamma$,
which has a 186 $\misec$ capture time. In addition,
``activity'' events are recorded for any event within the previous
51.2 $\misec$ and having
$> 17$ hit detector
phototubes or $> 5$ hit veto shield phototubes.

The first step in searching for $\bar \nu_e$ interactions is to
select electrons (the detector cannot distinguish between electrons and
 positrons)
with more than 300 hit
phototubes (highly efficient for energies above $28 \mev$), PID
information
consistent with a $\beta \sim 1$ particle,
$<2$ veto shield hits, and no ``activity'' events
in the previous 40 $\misec$.
The reconstructed position of the track midpoint is
required to be $> 35$ cm from the
locus of the
phototube faces.
Finally, events with three or more associated $\gamma$'s
are consistent with cosmic-ray neutrons and are eliminated.
The overall electron selection efficiency is $28 \pm 2 \%$.
In the $36<E_e<60$ MeV energy range, there are 135 such electron
events with the beam on and 1140 with the beam off, giving a beam-on
excess of $46.1\pm 11.9$ events.

The second step is to require  a correlated 2.2 MeV $\gamma$ with
a reconstructed distance, $\Delta r$, within 2.5 m of the electron,
a relative time, $\Delta t$, of less than 1 ms (imposed by the trigger),
and a number of hit phototubes, $N_\gamma$, between 21 and 50.
The efficiency for a neutron to be captured by a free proton and for the
2.2-MeV $\gamma$ to be found by these cuts is 63\%.
To determine if such a $\gamma$ is correlated with the electron or from an
accidental coincidence,
a function R of $\Delta r$, $\Delta t$, and $N_\gamma$ is defined
to be the ratio of approximate likelihoods for the two hypotheses.
Distributions of these quantities for
correlated $\gamma$'s are measured using cosmic ray neutron events.
We also compute the $\Delta r$ distribution with a Monte Carlo
simulation.
The R distributions for accidental $\gamma$'s are measured as
a function of electron position
using the large sample of electrons from cosmic-ray
muon decays.
The R distributions are shown in Fig. 1a, and Fig. 1b shows the
R spectrum for the beam-on minus beam-off data sample.

Requiring that a $\gamma$ be found with $R>30$
has an efficiency of 23\% for events with a recoil neutron and an
accidental rate of 0.6\% for events with no recoil neutron.
Fig. 2 shows the beam on minus beam off
energy distribution for events with
$R>30$. There are 9 beam-on and 17 beam-off events
between 36 and $60 \mev$, corresponding to a beam-on
excess of 7.7 events.
Table \ref{Sig}
lists the locations and energies for the 9 beam-on events.
When any of the electron selection criteria is relaxed, the
background increases slightly,
but the beam-on minus beam-off event excess
does not change significantly.

Table \ref{Back} lists the expected
number of background events in the $36<E_e<60$ MeV energy range for
$R>30$.
The {\sl beam-unrelated background\/} is
well determined from the thirteen-fold larger data sample collected
between accelerator pulses.
To set a limit on {\sl beam-related neutron
backgrounds}, events were selected which failed electron PID criteria
but were otherwise consistent with the correlated $e\gamma$
signature and in the electron energy range of interest.
The yield of beam-related neutron events of this type was $<3\%$ of
all neutrons when the beam was on.
Applying this ratio to neutrons {\sl passing\/} electron PID criteria,
the beam-related neutron background is bounded by 0.03 times the
total beam-unrelated background, and is thus negligible.
The largest {\sl neutrino background\/}, due to $\mu^-$ decay at rest
in the beam stop
followed by $\nuebar p \rightarrow e^+ n$ in the detector,
is calculated using the Monte Carlo beam simulation\cite{beam_mc}.
Another background with a recoil neutron arises from
$\bar \nu_{\mu} p \rightarrow \mu^+ n$ (including $\bar \nu_{\mu} C
\rightarrow \mu^+ n X$)
if the muon is lost (due to the ``activity'' threshold
or trigger inefficiency)
or if it is misidentified as an electron (e.g., if a fast decay made
the $\mu$ and $e$ look like a single particle).
This background is determined
from our measurement of $\nu_{\mu} C \rightarrow \mu^-
X$ \cite{Albert} and from our Monte Carlo detector simulation.\cite{MC}
Finally, the sum of all backgrounds involving {\sl accidental
$\gamma$'s}
is computed from the yield of electrons without correlated neutrons,
which
is measured using the likelihood fit described below.
The total estimated beam-related background for $R>30$ is thus
$0.79 \pm 0.12$ events,
which implies a net excess of 6.9 events in the
$36 <E_e < 60$ MeV energy range.
The probability that this excess is due to a statistical fluctuation is
$< 10^{-3}$.

While the $R>30$ sample demonstrates the existence
 of an excess, the size of the excess is better determined by utilizing
all electron data between 36 and $60\mev$.
The total numbers of beam-on and beam-off electron
events with correlated $\gamma$'s are obtained
from a likelihood fit to the R distributions at the electron positions.
The two ways of estimating
the $R$ distribution for correlated photons give
excesses of $18.3^{+9.5}_{-8.7}$
events (Monte Carlo method) and
$19.9^{+10.0}_{-9.1}$ events (cosmic neutron method).
Averaging these numbers and subtracting the neutrino background
with a neutron (2.7 events) gives an oscillation probability of
($0.34^{+0.20}_{-0.18}\pm 0.07$)\%,
where the first error is statistical and the second systematic.
The latter arises primarily from uncertainties in the neutrino
flux (7\%), effective fiducial volume (10\%), and $\gamma$
efficiencies (10\%).
The average of the fits implies that $27.0^{+8.9}_{-9.7}$ of the
beam-correlated electron events have no recoil neutron.
Background estimates from
$\nu_e ~^{12}C \rightarrow e^-~^{12}N$,
$\ \nu_e ~^{13}C \rightarrow e^-~^{13}N$, $\ \nu e \rightarrow
\nu e$, and other known neutrino interactions predict
$\sim 14$ events. \cite{LSNDref}

Cosmic-ray background is especially intense in the outer regions
of the detector and where the veto has gaps --
beneath the detector (low y),
and near the lower corner of the upstream end (low y and low z).
In an effort to find anomalous spatial concentrations of the
ocillation candidates,
we performed Kolmogorov tests on distributions
of various quantities, among which were y,
distance from the lower upstream corner, and
distance from the surface containing the photomultiplier faces.
These tests, done both with no photon criteria and with
$R>30$, gave probabilities above 25\% of consistency with
what is expected, with the
exception of one distribution not expected to be
sensitive to background; the distribution in x,
with no photon criteria, had a probability of 4\%.

We have also investigated alternative geometric criteria.
Removing the 5\% of the total volume having $y<-120$ cm and $z<0$
removes 32\% of the beam-off background, and results in
a net excess of $20.6^{+9.5}_{-8.7} \pm 4.1$ events,
corresponding to an oscillation probability of
($0.45^{+0.21}_{-0.19}\pm 0.10$)\%.
None of the $R>30$ events is in this area
of largest beam-off background.

The neutrino
oscillation probability for two-generation mixing can be expressed as
$P = (\sin^22\theta) \sin^2(1.27\Delta m^2 L/E)$,
where L is the distance (meters) between
the reconstructed positron position and the neutrino production
point and E is the neutrino energy (MeV) obtained from the measured
positron energy and direction.
A possible concern is the presence of $R>30$ events near and above 60 MeV.
But the Kolmogorov probability of consistency with a large $\Delta m^2$,
for example, oscillation hypothesis is 71\% for $36<E_e<60$ MeV and
13\% for $36<E_e<80$ MeV (ignoring any possible contribution from
decay-in-flight oscillation events).

If the observed excess is due to neutrino oscillations,
Fig. 3 shows the allowed region ($95\%$ C.L.)
of $\sin^22\theta$ vs. $\Delta m^2$ from
a maximum likelihood fit to the L/E distribution of
the 9 beam-on events in the $36<E_e< 60$ MeV energy range with $R>30$.
The result is renormalized to the measured oscillation probability of
0.34\% given above. The fit includes background subtraction,
smearing due to positron energy, position, and angular resolutions,
and the uncertainty of the neutrino production vertex.
The allowed region is not in conflict with
previous low energy decay-at-rest neutrino experiments E225 \cite{E225}
and E645 \cite{E645} at LAMPF. Some of the allowed region is excluded by
the ongoing KARMEN experiment\cite{KARMEN} at ISIS,
the E776 experiment at BNL\cite{wonyong},
and the Bugey reactor experiment \cite{BUGEY}.

In conclusion, the LSND experiment observes 9 electron events
in the $36 < E_e < 60$ MeV energy range which are correlated
in time and space with a low energy $\gamma$.
The total estimated background from
conventional processes is $2.1 \pm 0.3$ events,
so that the probability that the excess is
due to a statistical fluctuation is $< 10^{-3}$.
If the observed excess is interpreted as
$\bar \nu_{\mu} \rightarrow \bar \nu_e$ oscillations, it corresponds to
an oscillation probability of $0.34^{+0.20}_{-0.18} \pm 0.07\%$
for the allowed regions
shown in Fig. 3. If the excess is due to direct lepton number violation
and the spectrum of $\bar \nu_e$ is the same as for $\bar \nu_{\mu}$ in
$\mu^+$ decay, then the violation rate is the same as the above
oscillation probability. We plan to collect more data, and backgrounds
and detector performance continue to be studied.
These efforts are expected to improve the understanding of the
phenomena described here.

\paragraph*{Acknowledgements}

The authors gratefully acknowledge the support of Peter Barnes and
Cyrus Hoffman  during this work.  We also wish to thank the operating staff
at LAMPF for the magnificent performance of the accelerator.
The technical contributions of V.~Armijo, K.~Arndt, D.~Callahan,
B.~Daniel, S.~Delay, C.~Espinoza, C.~Gregory, D.~Hale, G.~Hart,
W.~Marterer, and  T.~N.~Thompson to the construction
and operation of LSND were invaluable. We also want to thank
 the students
M.~Albert, G.~Anderson, C.~Ausbrooks, B.~Bisbee, L.~Christofek,
M.~Davis, D.~Evans,  J.~George, J.~Hill, B.~Homann,
R.~Knoesel, S.~Lueder, S.~Mau, T.~Phan, F.~Schaefer,
M.~Strickland, B.J.~Vegara, M.~Volta,
S.~Weaver, and K.~Yaman for their help in making the
detector operational.

This work was supported by the U. S. Department of Energy and by
the National Science Foundation.

\begin{figure}
\caption{The distribution of R, the $\gamma$ likelihood parameter.
The leftmost bin corresponds to no $\gamma$ found within cuts (R=0),
properly normalized in area.
(a) Accidental photons (averaged over
the tank) and correlated photons (2 methods, described in text).
(b) Beam-on minus beam-off spectrum
for events in the $36<E_e<60$ MeV energy range. The dashed
histogram is the result of the R likelihood fit for events without a
recoil
neutron,
while the solid histogram is the total fit, including events with a
neutron.}
\end{figure}

\begin{figure}
\caption{
The electron energy distribution, beam-on minus beam-off, for
events with an associated 2.2 MeV $\gamma$ with $R>30$.
The dashed histogram shows the
expected background from known neutrino interactions.
The dotted curve is the expected distribution for
neutrino oscillations in the limit of large $\Delta m^2$,
normalized to the excess between 36 and 60 MeV.}
\end{figure}

\begin{figure}
\caption{The determination of $\sin^22\theta$ vs. $\Delta m^2$ from
a maximum likelihood fit to the L/E distribution of the 9 events which
satisfy the $R>30$ requirement, where $L/E$ is the neutrino distance to
energy ratio, normalized to the oscillation probability extracted from
the photon likelihood fit.
The shaded area is the allowed region (95\% C.L.)
from LSND. Not shown is the 20\% systematic uncertainty in the LSND
normalization. Also shown are 90\% C.L. limits from KARMEN (dotted
histogram), the BNL E776 experiment (dashed histogram), and
the Bugey reactor experiment (dot-dashed histogram).}
\end{figure}

\begin{table}
\caption{The
position, energy, and distance to the phototubes for the 9 beam-on
events
in the $36<E_e<60$ energy range with $R>30$. X, Y, and Z are the
lateral, vertical, and
longitudinal coordinates relative to the tank center.}
\label{Sig}
\begin{tabular}{lccccr}
Event&X(cm)&Y(cm)&Z(cm)&E(MeV)&D(cm)\\
\tableline

1&-66&-84&-77&47.8&115\\
2&56&-96&53&51.4&103\\
3&-36&196&-203&40.3&53\\
4&69&-146&153&44.3&53\\
5&-156&-79&-207&36.4&84\\
6&-221&-24&-309&56.9&36\\
7&-91&119&209&37.9&109\\
8&71&-99&-259&55.8&100\\
9&6&211&173&43.8&38\\
\tableline
\end{tabular}
\end{table}

\begin{table}
\caption{Expected
number of background events in the $36<E_e<60$ energy
range for $R>30$. The neutrinos are from either $\pi$ and $\mu$ decay
at rest (DAR) or decay in flight (DIF). Neutrino backgrounds with an
accidental
neutron signature are measured using the R likelihood fit described in
the text.}
\label{Back}
\begin{tabular}{lcr}
Background&Neutrino Source&Events with $R>30$\\
\tableline

Beam-unrelated&&$1.33\pm 0.32$\\
Beam-related n's &&$ <0.04$\\
$\bar \nu_e p \rightarrow e^+ n$&$\mu^- \rightarrow e^- \nu_{\mu} \bar
\nu_e$
DAR&$0.44\pm 0.06$\\
$\bar \nu_{\mu} p \rightarrow \mu^+ n$&$\pi^- \rightarrow
\mu^- \bar \nu_{\mu}$ DIF&$0.19\pm 0.08$\\
Accidentals &$\pi ,\mu$ DAR,DIF&$0.16\pm 0.06$\\
\tableline
Total &&$2.12 \pm 0.34$\\
\end{tabular}
\end{table}


\begin{references}
\bibitem{nuebkg}
The $\nuebar$ to $\numubar$ ratio is
given approximately by the product of the
$\pi^-/\pi^+$ ratio (0.2), the probability of $\pi^-$
decay in flight (0.03), and the probability of $\mu^-$ decay at rest
before capture in the Cu beam stop (0.12).
The $\nuebar$ spectrum is also softer than that of $\numubar$.
\end{references}
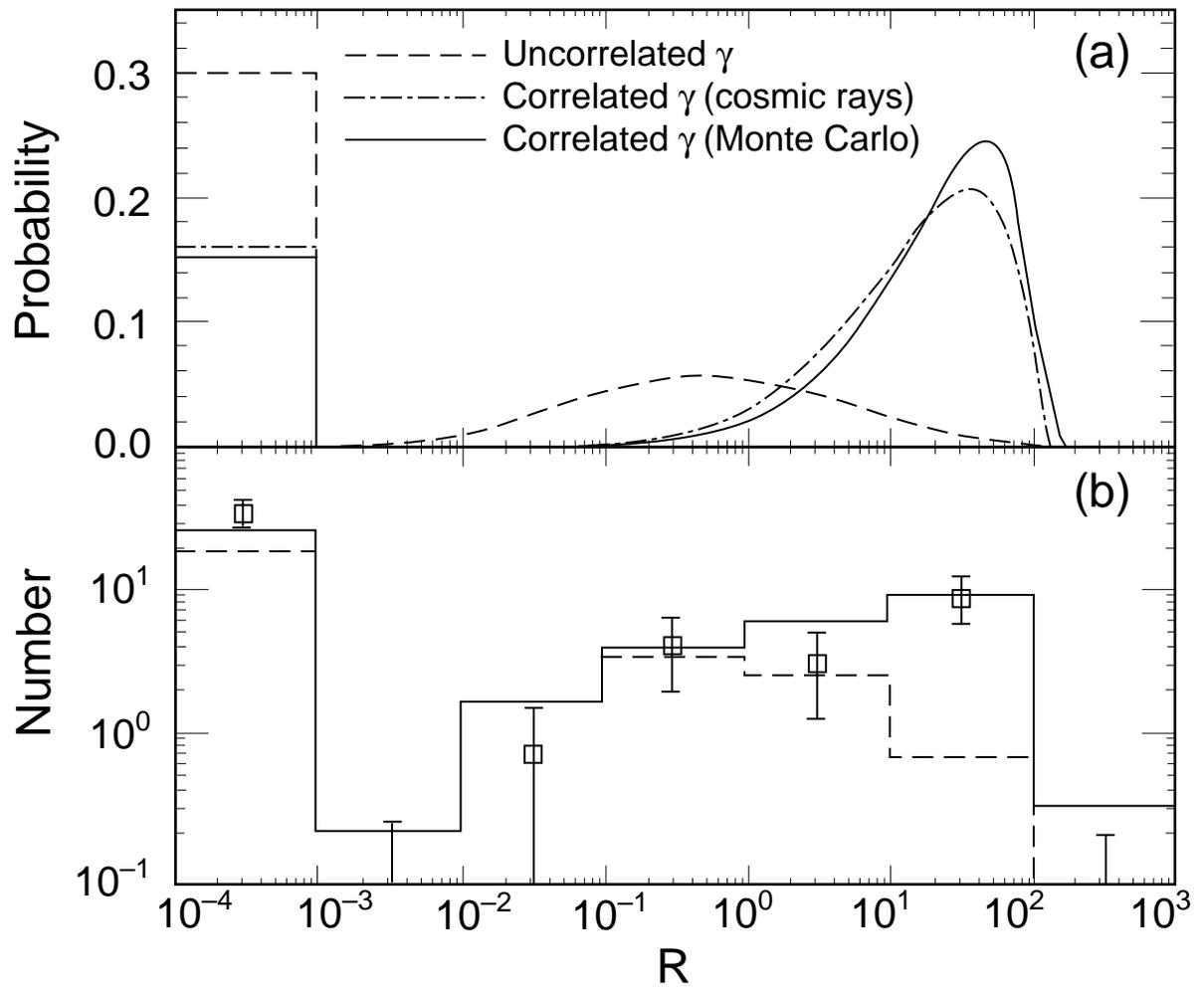
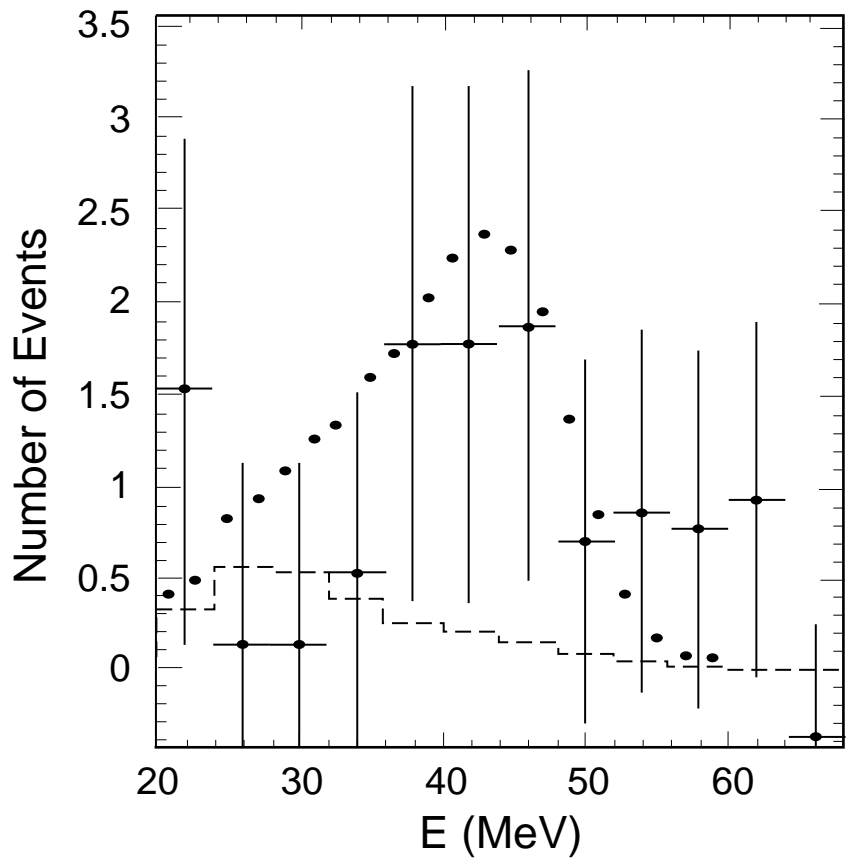
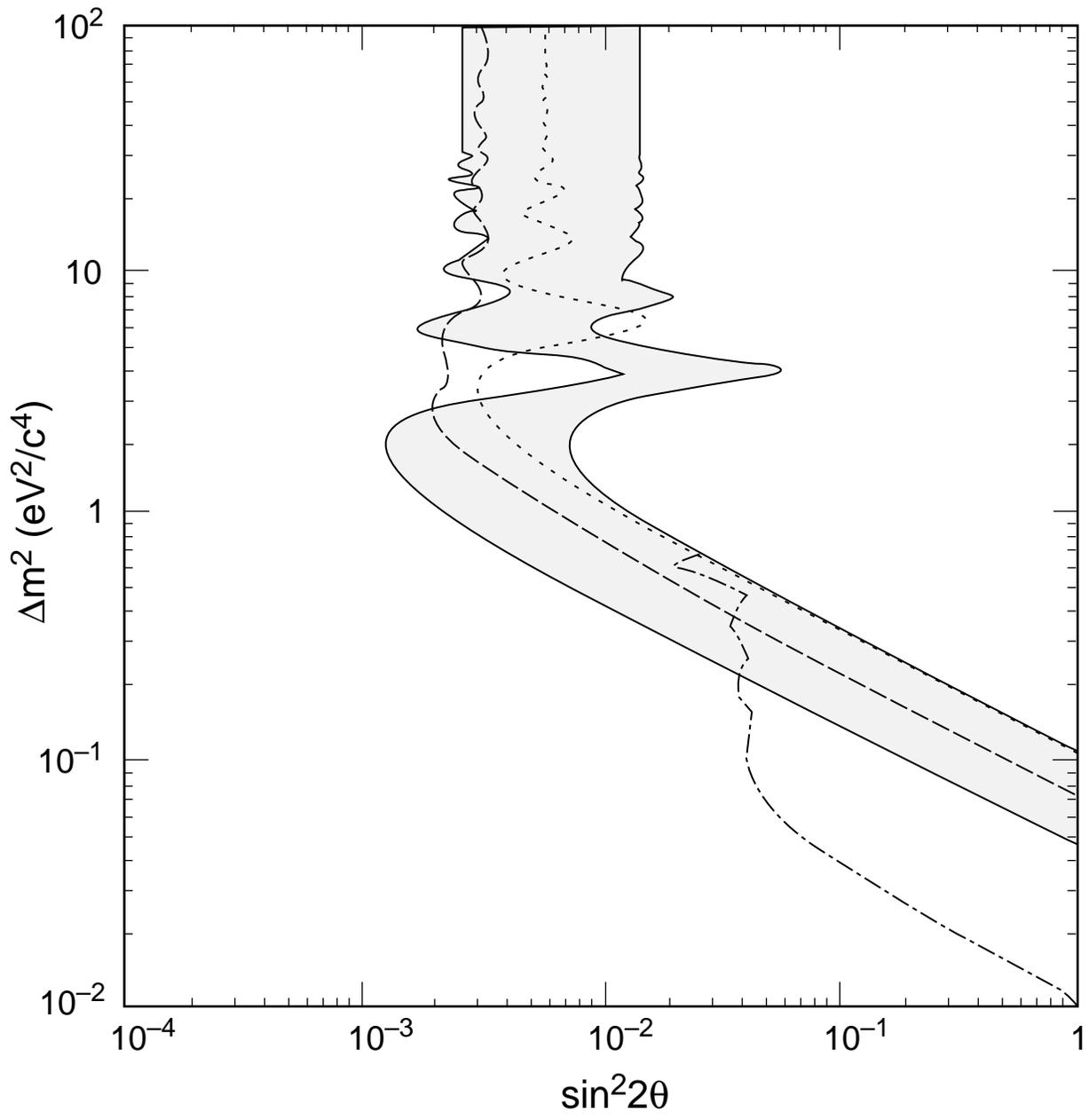
\end{document}